\documentclass[twocolumn, pra]{revtex4}
\usepackage[paperwidth=210mm,paperheight=297mm,centering,hmargin=2.0cm,vmargin=2.6cm]{geometry}
\usepackage[latin1]{inputenc}
\usepackage{amsmath}
\usepackage{amsfonts}
\usepackage{graphics}
\usepackage{epsfig}
\usepackage{dcolumn}

\newcommand{\beqa}{\begin{eqnarray}}
\newcommand{\eeqa}{\end{eqnarray}}

\newcommand{\ee}{\mathrm{e}}
\newcommand{\ii}{\mathrm{i}}

\newcommand{\bra}[1]{\langle #1 |}
\newcommand{\ket}[1]{|#1\rangle}
\newcommand{\braket}[2]{\langle #1 | #2 \rangle}

\newcommand{\beq}[1]{\begin{equation} \eqlab{#1}}
\newcommand{\eeq}{\end{equation}}
\newcommand{\bsub}{\begin{subequations}}
\newcommand{\esub}{\end{subequations}}
\newcommand{\nn}{\nonumber}

\newcommand{\eqlab}[1]{\label{eq:#1}}
\renewcommand{\eqref}[1]{Eq.~(\ref{eq:#1})}

\newcommand{\figref}[1]{Fig.~\ref{fig:#1}}

\newcommand{\dintegral}[4]{
\int_{#3}^{#4} \textrm{d}#2 \, #1
}

\usepackage{color}

\begin{document}

\title{Strong non-linearity-induced correlations for counter-propagating photons scattering on a two-level emitter}
\author{Anders Nysteen, Dara P. S. McCutcheon, and Jesper M{\o}rk}
\affiliation{DTU Fotonik, Department of Photonics Engineering, Technical University of Denmark, Building 343, 2800 Kgs. Lyngby, Denmark}

\date{\today}

\begin{abstract}
We analytically treat the scattering of two counter-propagating photons on a two-level emitter embedded in an optical waveguide. We find that the non-linearity of the emitter can give rise to significant pulse-dependent directional correlations in the scattered photonic state, which could be quantified via a reduction in coincident clicks in a Hong-Ou-Mandel measurement setup, analogous to a linear beam splitter. Changes to the spectra and phase of the scattered photons, however, would lead to reduced interference with other photons when implemented in a larger optical circuit. We introduce suitable fidelity measures which account for these changes, and find that high values can still be achieved 
even when accounting for all properties of the scattered photonic state. 

\end{abstract}

\maketitle

The realization of on-chip all-optical information processing requires the implementation of quantum computation schemes in 
integrated photonic circuits~\cite{Matthews09,Bruck15,Shadbolt12}, where the information is 
encoded in the travelling photons~\cite{Cirac97,Kimble08}. 
These schemes demand efficient single-photon sources~\cite{Claudon10,Chang06,Santori02}, photon detectors~\cite{Knill01,OBrien09}, 
and photon gates~\cite{Duan04,Tiecke14}. Since photons do not inherently interact, 
in order to realize two-photon gates such as the controlled phase or the CNOT 
gate~\cite{OBrien03,OBrien09,Zheng13}, optical non-linearities or post selection schemes are required. 
As Kerr-type non-linearities are usually weak in the few-photon limit, promising candidates to mediate the two-photon interactions 
are quantum mechanical two-level-systems. When two photons scatter off a two-level system, non-trivial correlations can be 
induced in the scattered state~\cite{Nysteen14,Fan10,Zheng10,Baragiola12,Roulet14,Valente12_NJP}, the nature of which ultimately 
determines the feasibility and scalability of optical circuits based on these components. 
Thus, a two-level-system embedded in a one-dimensional waveguide constitutes an important prototypical system in which to investigate 
few-photon scattering and non-linearity-induced photonic correlations. We note that such systems have been experimentally realized, 
for example, by self-assembled semiconductor quantum dots in photonic crystal waveguides, 
with emitter--waveguide coupling efficiencies reaching 
values in excess of 98\%~\cite{Arcari14}.

For two photons scattering on a single emitter, it is known that non-linearities are strongest 
when the photons are identical, and their spectral linewidths are comparable to that of the emitter, 
which results in the strongest correlations in 
the scattered photonic state~\cite{Nysteen14,Roulet14}. 
As such, identical input photons of a 
specific spectral lineshape are required. 
In the single photon case, however, it is know that such finite-width photons experience changes in both their 
spectra and phase as a result of the scattering process~\cite{Chen11,Fan10}, which is also known to be the case 
for two-photon scattering~\cite{Nysteen14}. Thus, once a photon has passed through an 
optical gate, it is no longer identical to an input photon, 
which in turn may limit the effectiveness of subsequent gates. 
This raises questions regarding the feasibility of integrating a large number of photonic gates needed to create complex optical circuits.
The purpose of this work is to explore how two-photon pulses are altered by the scattering process, and investigate 
how these alterations depend on the level of induced non-linearities. Interestingly, we find that non-linearaties can actually suppress spectral and phase changes, thereby increasing the similarity of the
scattered and input photons. As such, even when correctly accounting for all properties of the scattered photonic state, 
fidelities between the scattered and a desired directionally entangled state as high as $80\%$ can still be achieved.

\begin{figure}[b]
\includegraphics[width=0.45\textwidth]{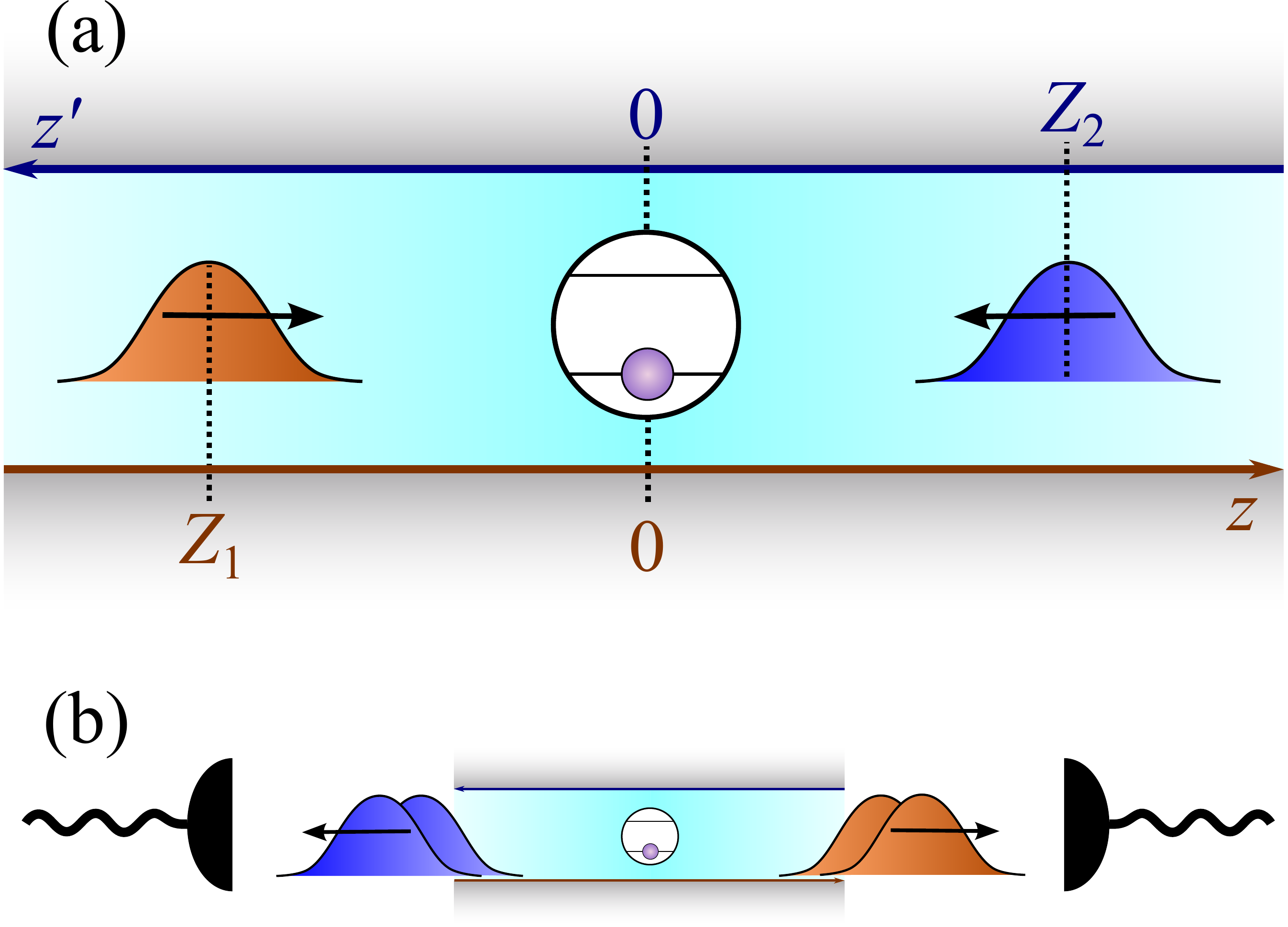}
\caption{\label{fig:sketch}(a) Two counter-propagating single-photon pulses propagate toward a two-level system in its ground state. 
(b) The post-scattering state is measured by detectors in each chiral waveguide mode sub-group.}
\end{figure}

Various methods have been used to analyze multi-photon scattering in systems consisting of a localized scattering 
object coupled to a waveguide. These include fully numerical approaches~\cite{Nysteen14,Moeferdt13}, 
as well as analytic approaches such as the input-output formalism~\cite{Fan10}, the real-space 
Bethe ansatz~\cite{Shen07}, the Lehmann-Symanzik-Zimmermann formalism~\cite{Shi11}, Laplace transforms~\cite{Huang13}, a wavefunction-based approach \cite{Valente12_NJP,Valente12_PRA},
and master equation formalisms~\citep{Baragiola12}. Several of these approaches allow for analytic determination of 
single- and two-photon scattering matrix elements, which directly relate the scattered state to the initial state of the system. 
Studies of two initially co-propagating photon pulses have been made for various scatterers coupling to waveguides, 
such as a single emitter~\cite{Fan10,Zheng10}, an emitter inside an optical cavitiy~\cite{Shi11,Shi13}, 
and a non-linear optical cavity~\cite{Liao10}.

Here we give a largely analytical description of the scattering of two counter-propagating photons 
impinging on a two-level emitter in a one-dimensional waveguide, as sketched in~\figref{sketch}(a). We use 
the scattering matrix formalism, and analyse the strong non-linearity-induced correlations in the scattered state 
for various input states. 
In addition, we introduce fidelity measures to 
quantify the induced correlation in the scattered state, 
taking both the spectrum and phase of the scattered state into account, and discuss their experimental interpretations. 

This paper is organized as follows: 
In Section~\ref{sec:general} we introduce our model. In Section~\ref{sec:onephot} we review the scattering of a single-photon pulse
on a two-level-emitter, and  introduce fidelity measures to quantify the similarity between the incoming and scattered photons. 
In Section~\ref{sec:twophot} 
the formalism is extended to the scattering of two counter-propagating single-photon pulses, where our fidelity measures 
are used to analyse induced correlations and spectral changes, and how these relate 
to the level of non-linearities.

\section{\label{sec:general}General theory}
We consider a quantum two-level-emitter coupled to two subsets of the modes in a one-dimensional waveguide as shown 
in Fig.~{\ref{fig:sketch}}, with the subsets differing by the direction of propagation. In the following we limit ourselves 
to lossless systems, though we note that this assumption could be relaxed by coupling our system to additional external 
reservoirs~\cite{Shen09_1,Rephaeli13}. Additionally, 
we neglect waveguide dispersion in the considered frequency interval, and we assume a localized scatterer (dipole approximation), 
i.e. that the scattering occurs only at a single point in space.

We describe the waveguide by two chiral mode subsets, being the right (mode index 1) and left (mode index 2) propagating modes. 
The corresponding Hamiltonian of the (bare) waveguide is 
\beqa
\tilde{H}_0 = \sum_{i=1}^2\dintegral{}{\tilde{k}}{0}{\infty}\hbar{\omega}(\tilde{k})c_i^\dagger(\tilde{k})c_i(\tilde{k}),
\eeqa 
where each mode in subsystem 1 and 2 is characterized by a wavevector $\smash{\tilde{k}}$, annihilation operator $\smash{c_i(\tilde{k})}$ and 
energy $\smash{\hbar{\omega}(\tilde{k})}$. In writing the Hamiltonian in this way we implicitly consider a single 
polarization of the waveguide modes. The Hamiltonian describing the emitter and its coupling to the waveguide is given by
\begin{equation}
\tilde{H}_1 = \frac{\hbar\omega_0}{2}\sigma_z\!+\!\hbar g\sum_{i=1}^2\int_0^{\infty}\!\!\!\mathrm{d}\tilde{k}[\sigma_+c_i(\tilde{k})+\sigma_-c_i^\dagger(\tilde{k})]
\end{equation}
where ${\omega}_0$ is the resonance frequency of the emitter and $g$ is its coupling strength 
to the waveguide, which is assumed to the frequency independent. This assumption is justified provided 
the linewidth of the emitter is small compared to the 
optical carrier frequencies of the photons. The operators $\sigma_+$ and $\sigma_-$ are the creation and annihilation operators 
of the emitter and $\sigma_z=\sigma_+\sigma_--\sigma_-\sigma_+$.

We consider pulses having the same carrier wavevector $k_p$ and corresponding frequency $\omega_p=\omega(k_p)$.
It is therefore convenient to work in a frame rotating with this carrier frequency~\cite{Fan10}. 
We relate the frequencies of the waveguide modes $\omega(\tilde{k})$ to their wavevectors using a 
Taylor expansion around $k_p$, giving
\beqa
\omega(\tilde{k}) \approx \omega_p+v_g(\tilde{k}-k_p)
\eeqa
with $\smash{v_g=(\partial \omega/\partial \tilde{k})|_{\tilde{k}=k_p}}$ being the group velocity. The rotating frame is defined 
by the transformation $\tilde{H}\to H=U\tilde{H}U^{\dagger}+\ii\hbar(\partial_t U^{\dagger})U$ with 
$U=\exp[\ii\omega_p t(\sigma_z/2+\sum_{i}\int_0^{\infty}\mathrm{d}\tilde{k} c_i^{\dagger}(\tilde{k})c_i (\tilde{k}))]$. We find 
$H=H_0+H_1$ with 
\begin{equation}
H_0=\hbar v_g\sum_{i=1}^2\int_{-\infty}^{\infty}\mathrm{d} k \,k\, a_i^{\dagger}(k)a_i(k),
\label{H0}
\end{equation}
where $k=\tilde{k}-k_p$ and we have defined the new annihilation operators $a_i(k)=c_i(k+k_p)$. 
The interaction Hamiltonian is now given by 
\begin{equation}
H_1 = \frac{\hbar\Delta}{2}\sigma_z+\hbar g\sum_{i=1}^2\int_{-\infty}^{\infty}\!\!\mathrm{d}k[\sigma_+a_i(k)+\sigma_-a_i^\dagger(k)],
\label{Ham_trans}
\end{equation}
with $\Delta={\omega}_0-k_pv_g$ the detuning of the carrier frequency from the emitter transition frequency. We note 
that in obtaining Eqs.~({\ref{H0}}) and ({\ref{Ham_trans}}) we have extended the lower limits of 
integration from $-k_p$ to $-\infty$. This approximation is justified since we will be interested in pulses with wavevectors centered around 
$k=\tilde{k}-k_p=0$ and whose widths are much smaller than $k_p$.

\section{\label{sec:onephot}Single-photon scattering}
Before we consider the scattering of two photons, we first review the single-photon scattering case and introduce the scattering matrix 
formalism. We consider cases for which the emitter is initially in its ground state. Following Fan et al.~\cite{Fan10}, 
we relate the state of the system long after the scattering process to the initial state through the scattering matrix $S$. 
For non-linear scatterers the scattering matrix will in general be frequency dependent, and for a localized scatterer
the scattering elements are defined as~\cite{Fan10}
\beqa
_1\bra{p}S^{(1)}\ket{k}_1 = {}_2\bra{p}S^{(1)}\ket{k}_2 = \bar{t}_k\delta(p-k), \\
_2\bra{p}S^{(1)}\ket{k}_1 = {}_1\bra{p}S^{(1)}\ket{k}_2 = \bar{r}_k\delta(p-k),
\eeqa
where the notation implies, $a_i^\dagger(k)\ket{\phi}=\ket{k}_i$ with $\ket{\phi}$ the vacuum, and 
where $\bar{t}_k$ and $\bar{r}_k$ are the frequency-dependent single-photon transmission and reflection coefficients, respectively. 
The delta-functions reflect momentum conservation, and as no external loss channels are present, $|\bar{t}_k|^2+|\bar{r}_k|^2=1$. 

An arbitrary single photon state propagating to the right is written
\beqa
\ket{\xi_0}=\dintegral{}{k}{-\infty}{\infty}\xi(k)a_1^\dagger(k)\ket{\phi}, \eqlab{singlephot_input1}
\eeqa
where $\xi(k)$ is the wavepacket in momentum space. 
The post-scattering state corresponding to the incoming state expressed in~\eqref{singlephot_input1} is 
defined as $\ket{\xi}_{t\to\infty}=S^{(1)}\ket{\xi_0}$. It is obtained by inserting the  
identity operator 
\beqa
\openone = \sum_{i=1,2}\dintegral{}{p}{-\infty}{\infty}\ket{p}_i{}_i\bra{p},
\eeqa
from which we find 
\begin{equation}
\ket{\xi}_{t\rightarrow\infty}
=\dintegral{}{p}{-\infty}{\infty}\bar{t}_p\xi(p)\ket{p}_1+\dintegral{}{p}{-\infty}{\infty}\bar{r}_p\xi(p)\ket{p}_2.
\end{equation}
The two terms above reflect the fact that the photon can be transmitted or reflected. 
The scattering probabilities are defined as 
$P_i={_{t\rightarrow\infty}}\bra{\xi}\dintegral{}{p}{-\infty}{\infty}\ket{p}_i{}_i\braket{p}{\xi}_{t\rightarrow\infty}$, 
with the transmission and reflection probabilities corresponding to $i=1$ and $i=2$ respectively. 
These probabilities are found to be,
\begin{equation}
P_1=\dintegral{}{p}{-\infty}{\infty}|\bar{t}_p\xi(p)|^2, 
\qquad
P_2=\dintegral{}{p}{-\infty}{\infty}|\bar{r}_p\xi(p)|^2.
\end{equation}

The theory above applies to any localized scatterer interacting with two chiral waveguide modes. 
We now specifically consider the emitter-waveguide system sketched in~\figref{sketch} and described by the 
Hamiltonians in Eqs.~({\ref{H0}}) and (\ref{Ham_trans}). In this system the reflection and transmission coefficients may be found 
through calculation of the single-photon scattering matrix elements~\cite{Fan10}, which gives 
\begin{equation}
\bar{t}_k = \frac{k-\Delta}{ k-\Delta +\ii\Gamma/(2v_g)},
\qquad
\bar{r}_k = \frac{-\ii\Gamma/(2v_g)}{ k-\Delta +\ii\Gamma/(2v_g)}, \eqlab{one_reftrans}
\end{equation}
where $\Gamma=4 \pi g^2/v_g$ is the decay rate of the emitter. 
We note that this form of $\Gamma$ can be found through Fermi's Golden Rule, and is valid for a lossless 
system in which the emitter couples equally to modes propagating in both directions.

\begin{figure}
\includegraphics[scale=0.65]{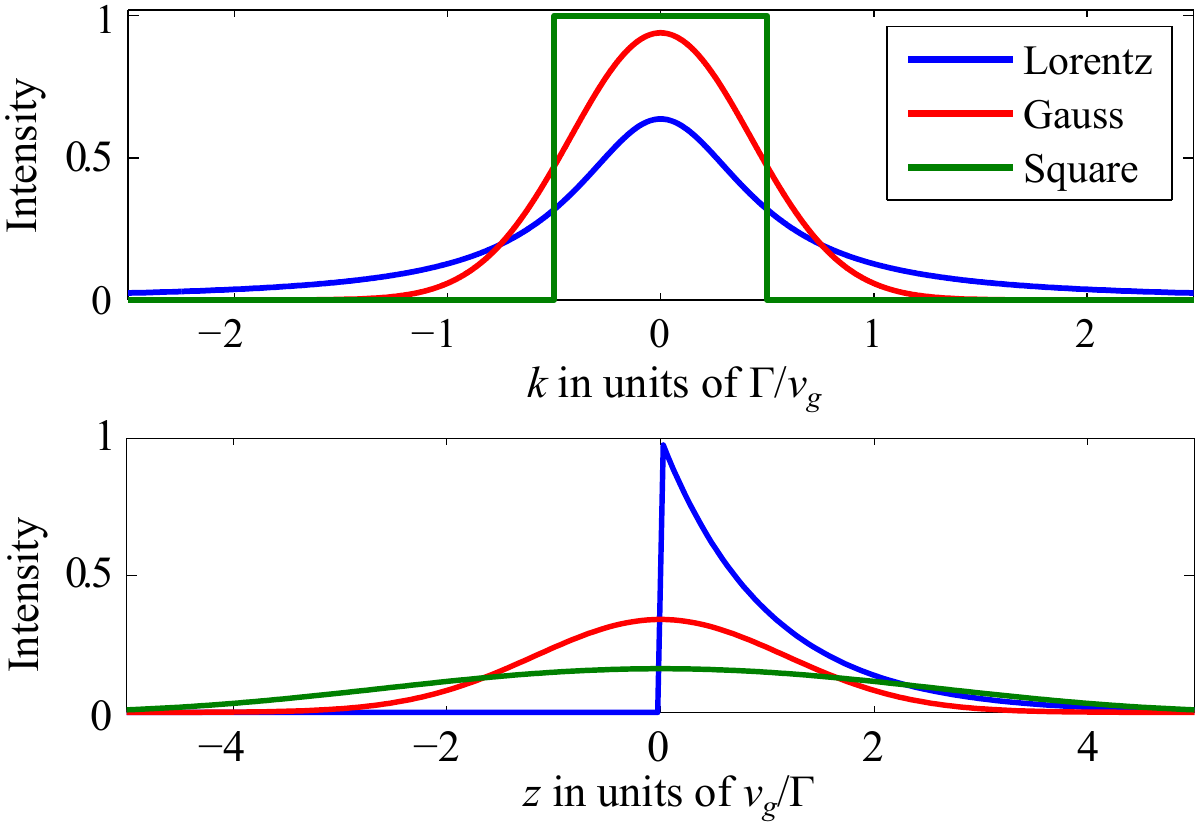}
\caption{\label{fig:single_shape} (a) Intensity spectra, $|\xi(k)|^2$, for the three single-photon 
wavepackets in Eqs.~(\ref{eq:puls1})-(\ref{eq:puls3}), here plotted with $\sigma=\Gamma/v_g$. 
(b) Corresponding spatial pulse profiles, $|\xi(z)|^2$, with large values of $z$ corresponding to the front part of the pulse, arriving first at the position of the emitter.}
\end{figure}

We consider three transform-limited incoming pulse shapes, i.e. Lorentzian, Gaussian, and step-function~\cite{Gulati14,Keller04}, which 
are defined by the wavepackets 
\beqa
\xi_\text{Lor}(k)&=& \frac{\sqrt{\sigma/(2\pi)}}{k-\ii \sigma/2}, \eqlab{puls1}\\
\xi_\text{Gauss}(k)&=& (\pi\sigma'^2)^{-1/4}\ee^{-k^2/(2\sigma'^2)}, \\
\xi_\text{square}(k)&=& \sigma^{-1/2}\theta(\sigma/2-|k|), \eqlab{puls3}
\eeqa
with $\sigma'=(2\sqrt{\ln(2)})^{-1}\sigma$ for the Gaussian wavepacket, and where $\theta(k)$ is the Heaviside step function. 
All wave packets are normalised such that $\dintegral{}{k}{-\infty}{\infty}|\xi(k)|^2=1$, and have a spectral full 
width--half maximum of $\sigma$. 
The intensity spectra of the pulses are shown in~\figref{single_shape} together with the spatial pulse profiles, 
defined as the Fourier transform $\xi(z)=(2\pi)^{-1/2}\dintegral{}{k}{-\infty}{\infty}\xi(k)\exp[\ii k z]$.

\begin{figure}
\includegraphics[scale=0.6]{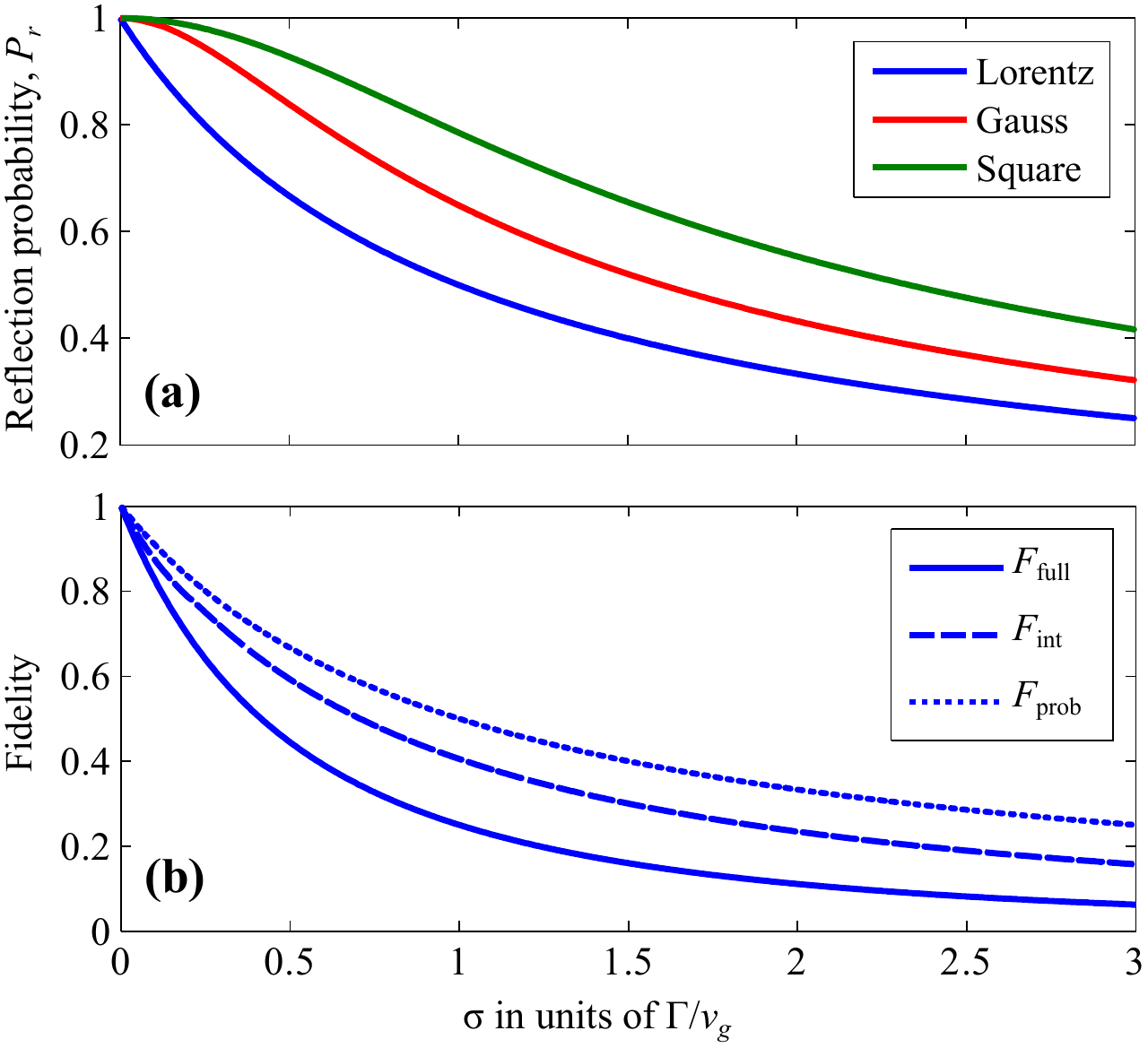}
\caption{\label{fig:single_prob} (a) Reflection probabilities as a function of pulse width for the three pulse shapes given in 
Eqs.~(\ref{eq:puls1})-(\ref{eq:puls3}) with carrier frequencies resonant with the emitter. (b) The three fidelity measures 
from Eqs. (\ref{eq:single_fid1})-(\ref{eq:single_fid3}) for scattering of a 
Lorentzian pulse shape.}
\end{figure}

A comparison of the resulting reflection probabilities is shown in~\figref{single_prob}(a) for each of the three single-photon 
wavepackets in Eqs. (\ref{eq:puls1})-(\ref{eq:puls3}), as has been calculated in previous works~\cite{Chen11,Zheng10}. 
The frequency components of the pulse closest to the transition energy of the emitter interact most strongly, 
and those at the exact frequency of the emitter ($k=0$) are perfectly reflected~\cite{Chen11}. Thus, as the spectral 
pulse width is decreased, the reflection probability increases, and reaches unity for resonant monochromatic 
pulses ($\sigma\to0$). In the opposite limit of $\sigma\rightarrow\infty$, only a vanishing fraction of frequency components 
overlap with the spectrum of the emitter, resulting in complete transmission since the pulse does not interact with the emitter. 
The pulse shape also has an important impact on the reflection probability. 
Since the Lorentzian has the largest spread of frequency components for a given FWHM, 
it interacts least with the emitter and correspondingly results in the lowest reflection probability.

If many emitters are to be implemented in a larger sequence of photonic devices or gates, it is important that 
scattered photons maintain their spectral properties, i.e. the pulse shape and phase variation across the pulse. 
We therefore seek to define measures to compare the scattered state with some desired output state. We define 
three measures for this purpose, each with a different physical significance. We use the quantum state fidelity~\cite{Kokbook} 
as a measure of the degree to which the scattered state is quantum mechanically identical to 
the desired state (neglecting overall phase differences), 
\beqa
F_{\mathrm{full}} &=& |\bra{\xi_\text{des}}\xi\rangle_{t\rightarrow \infty}|^2 \eqlab{F_exp},
\eeqa
where $\ket{\xi_{\mathrm{des}}}$ is the desired state. 
When the desired state is equal to the input state but propagating in the opposite direction (i.e. a reflected but otherwise unchanged state,  
$\ket{\xi_{\mathrm{des}}}=\int_{-\infty}^{\infty}\xi(p)\ket{p}_2$), we find 
\beqa
F_\text{full} = \bigg| \dintegral{}{p}{-\infty}{\infty} \bar{r}(p)|\xi(p)|^2 \bigg|^2. \eqlab{single_fid1}
\eeqa
This fidelity measure is appropriate, for example, if the scattered photon were to interfere with a second input photon 
in a Hong--Ou--Mandel-type interference experiment 
(where two truly indistinguishable photons exit a 50/50 beam-splitter in the same arm).

For cases in which the phase of the scattered state is unimportant, but instead we are interested in how the intensity in the pulse is 
distributed spectrally, the similarity between the scattered and the desired pulse may be characterized by 
\beqa
F_\text{int} = \left( \dintegral{}{p}{-\infty}{\infty} |\bar{r}(p)||\xi(p)|^2 \right)^2 .
\eeqa
This fidelity measure would be relevant when comparing the energy distributions in the scattered 
and desired pulse, which could be achieved by introducing spectrometers in a setup as sketched in~\figref{sketch}(b), 
but disregarding the arrival times at the detectors. 
It can be seen to be the limiting form of a spatial definition of a fidelity measure, defined as 
\beqa
F_\textrm{spat} = \max_{\delta z}\bigg| \dintegral{}{z}{-\infty}{\infty}\xi^*(z)\xi_\textrm{scat}(z-\delta z) \bigg|^2,
\eeqa
where $\xi_\textrm{scat}(z)$ is the spatial representation of the scattered wavepacket, which after the 
scattering may be displaced by $\delta z$ in the rotating frame due to a delay caused by the absorption in the emitter. 
Using the defined Fourier transform and H\"older's inequality, we find $F_\text{int}$ is an upper bound for this spatial fidelity, 
\begin{equation}
F_\textrm{spat} = \max_{\delta z}\bigg| \dintegral{}{p}{-\infty}{\infty}|\xi(p)|^2\bar{r}(p)\ee^{\ii\delta z k} \bigg|^2 \leq F_\text{int}.
\end{equation}

Finally, when neither the phase nor the spectral distribution are important, the scattered state may be projected onto a basis 
which merely counts the number of photons in each waveguide mode, e.g. as in \figref{sketch}(b). The fidelity in this case simply becomes 
the probability of detecting a photon in the desired output mode (here the reflected field),
\beqa
F_\text{prob}=P_r. \eqlab{single_fid3}
\eeqa

We evaluate these fidelities for a Lorentzian input, and show the results in~\figref{single_prob}(b). For 
$F_\textrm{full}$ and $F_\textrm{prob}$ we find the exact expressions 
\beqa
F_\text{full} &=& \frac{(\tilde{\Gamma})^2}{(\tilde{\Gamma}+\sigma)^2+4\Delta^2}, \\
F_\text{prob} &=& \frac{(\tilde{\Gamma}+\sigma)\tilde{\Gamma}}{(\tilde{\Gamma}+\sigma)^2+4\Delta^2},
\eeqa
where $\tilde{\Gamma}=\Gamma/v_g$. 
From~\figref{single_prob}(b) we see that $0\leq F_\text{full} \leq F_\text{int} \leq F_\text{prob} \leq 1$. This 
reflects the progressively less stringent criteria of these three measures. 
As the desired state in each case is a fully reflected state, the fidelities are largest for small FWHMs.

\section{\label{sec:twophot}Two-photon scattering}
We now turn to the main focus of this work, and extend our formalism to describe the scattering of two-photon states. 
In the single-photon case, energy conservation implied that an approximately monochromatic single-photon wavepacket 
would scatter without changing its frequency. In the two-photon case, energy conservation only demands 
that the \emph{sum} of the energies of the two incoming and two scattered photons is conserved. 
According to Fan et al.~\cite{Fan10}, we can define a two-photon scattering matrix, $S^{(2)}$, in a 
similar way to $S^{(1)}$, which contains terms describing single-photon scattering, and also additional terms  
stemming from two-photon scattering processes. The additional terms involve four-wave mixing mechanisms 
between the two incoming and two scattered photons~\cite{Fan10,Nysteen14}.

In the rotating frame, a general two-photon state in the momentum representation is written
\begin{align}
\ket{\beta}=\textstyle{\frac{1}{\sqrt{2}}}&\dintegral{}{k}{-\infty}{\infty}\dintegral{}{k'}{-\infty}{\infty}\beta_{11}(k,k')a_1^\dagger(k)a_1^\dagger(k')\ket{\phi} \nn \\ 
+\textstyle{\frac{1}{\sqrt{2}}}&\dintegral{}{k}{-\infty}{\infty}\dintegral{}{k'}{-\infty}{\infty}\beta_{22}(k,k')a_2^\dagger(k)a_2^\dagger(k')\ket{\phi}\nn\\
+ &\dintegral{}{k}{-\infty}{\infty}\dintegral{}{k'}{-\infty}{\infty}\beta_{12}(k,k')a_1^\dagger(k)a_2^\dagger(k')\ket{\phi}, \eqlab{twostate}
\end{align}
normalized such that 
$\dintegral{}{k}{-\infty}{\infty}\dintegral{}{k'}{-\infty}{\infty}(|\beta_{11}(k,k')|^2+|\beta_{12}(k,k')|^2+|\beta_{22}(k,k')|^2)=1$. 
Introducing the notation $\ket{kk'}_{ii'} = \left\{{}_{i'i}\bra{k'k}\right\}^\dagger = a_i^\dagger(k)a_{i'}^\dagger(k')\ket{\phi}$, 
the two-photon scattering elements are~\cite{Fan10} 
\begin{align}
&{}_{jj'}\bra{pp'}S^{(2)}\ket{kk'}_{ii'} = \alpha_{ji,k}\alpha_{j'i',k'}\delta(k-p)\delta(k'-p') \nn\\
&\quad\quad\quad\quad\quad\quad\quad\quad+\alpha_{j'i,k}\alpha_{ji',k'}\delta(k-p')\delta(k'-p) \nn\\
&\quad\quad\quad\quad\quad\quad\quad\quad+\textstyle{\frac{1}{4}}B_{pp'kk'}\delta(p+p'-k-k'),
\end{align}
for $i,j\in\{1,2\}$, and where
\beqa
\alpha_{ji,k} = \begin{cases} \bar{t}_k & 
\text{if } i=j \\
\bar{r}_k   & \text{if } i\neq j   \end{cases}
\eeqa
are the single photon reflection and transmission matrix elements. 
Here $B_{pp'kk'}$ describes interactions between the two incoming and the two scattered photons, and 
is determined by the specific localized scatterer considered. 
As in the single-photon case, to find the scattered state, we insert the identity 
operator, which is now given by
\begin{align}
&\openone=\dintegral{}{p}{-\infty}{\infty}\dintegral{}{p'}{-\infty}{\infty}\nn\\
&\bigg[\frac{1}{2}\ket{p'p}_{11}{_{11}}\bra{pp'}
+\ket{p'p}_{21}{_{12}}\bra{pp'}+\frac{1}{2}\ket{p'p}_{22}{_{22}}\bra{pp'}\bigg].
\end{align}
If we assume an initial state consisting of two counter-propagating photons, $\beta_{12}$ is the 
only non-zero expansion coefficient in~\eqref{twostate}, which results in the post-scattering state
\begin{align}
&\ket{\beta}_{t\rightarrow\infty}=\dintegral{}{p}{-\infty}{\infty}\dintegral{}{p'}{-\infty}{\infty}\bigg\{ \nn\\
\frac{1}{2}&\bigg[(\bar{t}_{p}\bar{r}_{p'}+\bar{r}_{p}\bar{t}_{p'})\beta_{12}(p,p')+\textstyle{\frac{1}{4}}b_{12}(p,p') \bigg]a_1^\dagger(p)a_{1}^\dagger(p')\nn \\
+&\bigg[(\bar{r}_{p}\bar{r}_{p'}+\bar{t}_{p}\bar{t}_{p'})\beta_{12}(p,p')+\frac{1}{4}b_{12}(p,p') \bigg]a_1^\dagger(p)a_{2}^\dagger(p') \nn \\
+\frac{1}{2}&\bigg[(\bar{t}_{p}\bar{r}_{p'}+\bar{r}_{p}\bar{t}_{p'})\beta_{12}(p,p')+\frac{1}{4}b_{12}(p,p') \bigg]a_2^\dagger(p)a_{2}^\dagger(p') \nn\\
&\hskip5.5cm \bigg\}\ket{\phi}. \eqlab{two_long_scat_term2}
\end{align}
The first term in each of the three square brackets in~\eqref{two_long_scat_term2} represents single photon scattering 
processes, where the first factor contains the appropriate combinations of transmission and reflection coefficients connecting 
the initial and final photon configurations. Multi-photon process are contained in the pulse-dependent contribution $b_{12}(p,p')$, 
which describes processes induced by the emitter non-linearity, and is given by
\begin{equation}
b_{12}(p,p')=\dintegral{}{k}{-\infty}{\infty}\beta_{12}(k,p+p'-k)B_{pp'k(p+p'-k)}.
\eqlab{Fan_b}
\end{equation}

We define $P_{11}$ ($P_{22}$) as the probability that both photons are measured propagating in waveguide mode 
1 (mode 2), and $P_{12}$ the probability that one photon propagates in each waveguide mode. 
From~\eqref{two_long_scat_term2} we find 
\begin{align}
P_{11}&=\frac{1}{2}\dintegral{}{p}{-\infty}{\infty}\dintegral{}{p'}{-\infty}{\infty}\nn\\
&\bigg|(\bar{t}_{p}\bar{r}_{p'}+\bar{r}_{p}\bar{t}_{p'})\beta_{12}(p,p')+\frac{1}{4}b_{12}(p,p') \bigg|^2, \eqlab{eq_P11}\\
P_{12}&=\dintegral{}{p}{-\infty}{\infty}\dintegral{}{p'}{-\infty}{\infty} \nn\\
&\bigg|(\bar{t}_{p}\bar{t}_{p'}+\bar{r}_{p}\bar{r}_{p'})\beta_{12}(p,p') +\frac{1}{4}b_{12}(p,p') \bigg|^2,
\end{align}
with $P_{11}=P_{22}$ and $P_{11}+P_{12}+P_{22}=1$. 

As in the single photon case, we will be interested in comparing the scattered two-photon state described 
by~\eqref{two_long_scat_term2} to some desired state using the fidelity measures we have 
introduced in Eqs. (\ref{eq:single_fid1})-(\ref{eq:single_fid3}).
For this purpose we consider the scattered two-photon state that would be obtained if the scatterer were replaced by a perfect 50/50 
beam-splitter, which preserves the shape and phase of the input photons. In this case 
the scattered state would be 
\begin{align}
\ket{\beta_\textrm{des}}&=
\frac{1}{\sqrt{2}}\dintegral{}{k}{-\infty}{\infty}\dintegral{}{k'}{-\infty}{\infty}\beta_{12}(k,k')\nn\\
&\times\bigg[\frac{1}{\sqrt{2}}a_1^\dagger(k)a_1^\dagger(k')+\frac{1}{\sqrt{2}}a_2^\dagger(k)a_2^\dagger(k')\bigg]\ket{\phi}.
\end{align} 
With this desired state our fidelity measures become 
\begin{align}
\!\!F_\text{full} &= \!\bigg| \!\int_{-\infty}^{\infty}\!\!\!\mathrm{d}p\int_{-\infty}^{\infty}\!\!\!\mathrm{d}p'
\beta_{12}^*(p,p')\beta_{12}'(p,p')
 \bigg|^2 \eqlab{two_2fid1}\\
\!\!F_\text{int} &= \!\bigg[\!\int_{-\infty}^{\infty}\!\!\!\mathrm{d}p\int_{-\infty}^{\infty}\!\!\!\mathrm{d}p'|\beta_{12}(p,p')|\,|\beta_{12}'(p,p')|\bigg]^2 \\
F_\text{prob} &= P_{11}+P_{22} \eqlab{two_2fid3}
\end{align}
with $\beta_{12}'(p,p')=(\bar{t}_{p}\bar{r}_{p'}+\bar{r}_{p}\bar{t}_{p'})\beta_{12}(p,p')+\frac{1}{4}b_{12}(p,p')$. 

\subsection{Two-level scatterer}
The theory presented above is valid for any localized scatterer. 
We now specifically consider the two-level-emitter--waveguide system described by Eq.~(\ref{Ham_trans}), 
and focus here only on pulses starting equidistantly from the emitter. 
Furthermore, we only treat pairs of input pulses with the same spectral linewidth, 
although the formalism can be straightforwardly extended to more general cases. For a two-level-emitter 
the single photon transmission and reflection matrix elements $\bar{t}_k$ and $\bar{r}_k$ are given by~\eqref{one_reftrans}, 
while the two-photon scattering element is~\cite{Fan10}
\beqa
B_{pp'kk'}=\ii\frac{\sqrt{\Gamma}}{\pi}s_{p}s_{p'}(s_k+s_{k'}),
\eeqa
where
\beqa
s_k = \frac{\sqrt{\Gamma}/v_g}{k-\Delta+\ii\Gamma/(2v_g)}.
\eeqa
We only consider uncorrelated photon input states, and as such $\beta(k,k')$ is a symmetrized product of two 
single-photon wavepackets $\beta(k,k')=[\xi(k)\xi'(k')+\xi'(k)\xi(k')]$, which is normalized as 
$\dintegral{}{k}{-\infty}{\infty}\dintegral{}{k'}{-\infty}{\infty}|\beta(k,k')|^2=1$. 

\begin{figure}
\includegraphics[scale=0.6]{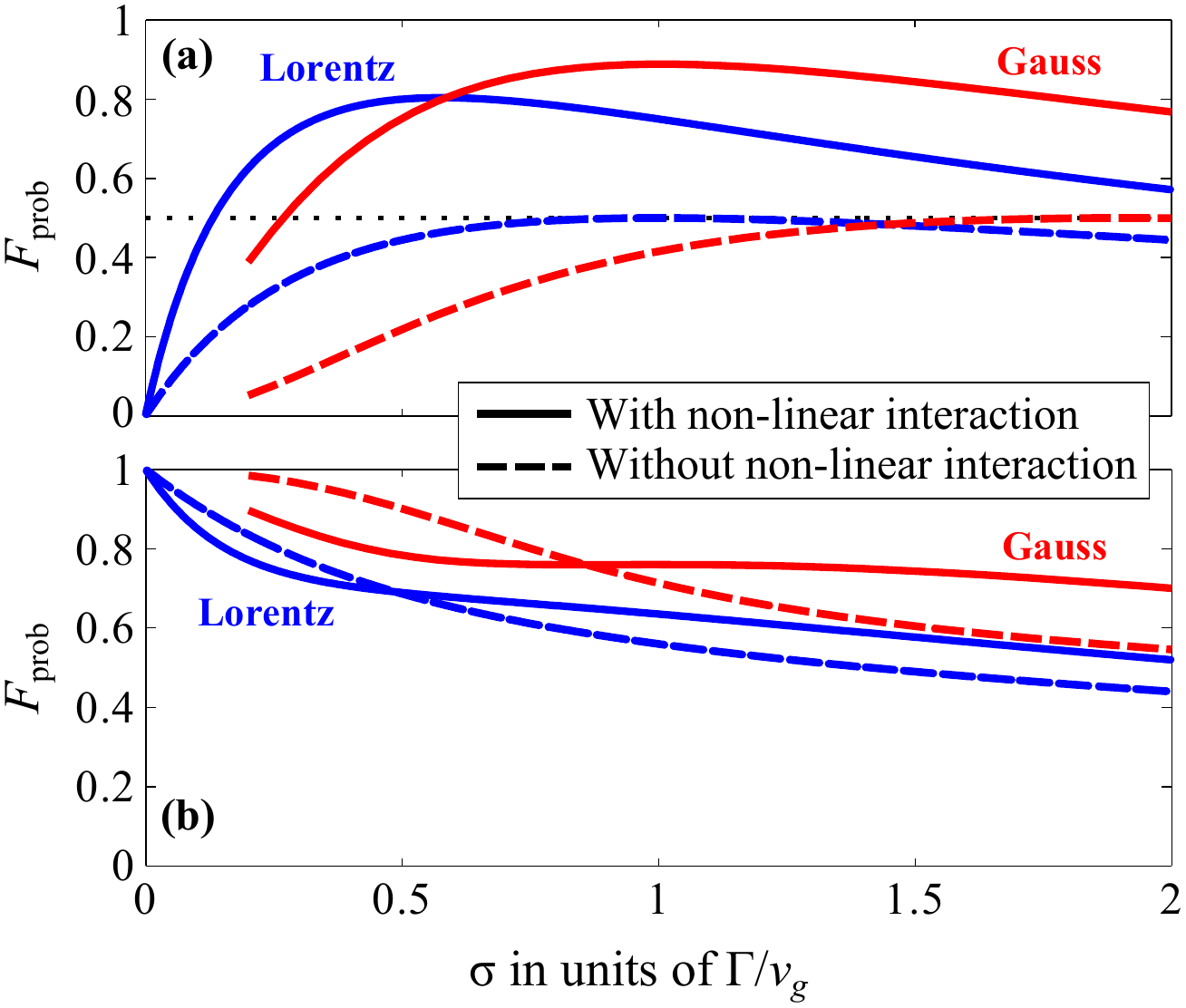}
\caption{\label{fig:comparison} Degree of directional entanglement $F_\text{prob}$ plotted for 
varying incoming pulse widths, shown for both Lorentzian and Gaussian input pulses (solid lines). 
The corresponding values obtained when neglecting two-photon scattering terms are also shown (dashed lines). 
(a) Pulse and emitter are resonant, $\Delta=0$. (b) Pulse and emitter detuned by $\Delta = \Gamma/(2v_g)$.}
\end{figure}

We begin our analysis of the scattered state by considering correlations in 
photon detection events in the two waveguide mode subsets, as depicted in \figref{sketch}(b). 
In this case no information regarding the spectrum and phase of the scattered photons is obtained, and 
the appropriate fidelity measure is $F_\text{prob}$, which is equal to $1$ minus the 
probability of detecting a coincidence in the two detectors, i.e. for $F_\text{prob}=1$ no 
coincidence events are measured (a perfect Hong-Ou-Mandel dip would be observed).  
In \figref{comparison}(a), $F_\text{prob}$ is calculated for Gaussian and Lorentzian input pulses for zero detuning $(\Delta=0)$, 
for which the interaction between the pulses and emitter is greatest. We see that very high fidelities are obtained, reaching 
values of $\sim 80\%$ for the Lorentzian input and $\sim 90\%$ for the Gaussian. 
Maximal correlations are achieved in the 
regime where the emitter and pulse linewidth are similar, as has been demonstrated numerically in earlier work~\cite{Nysteen14}. 
Interestingly, although the Lorentzian pulse shape is well-known to be the optimal pulse shape for maximally exciting the 
two-level emitter with a single photon~\cite{Rephaeli10}, it is not the optimal shape for maximizing the directional correlations in the scattered state.

The high fidelities obtained demonstrate that the scattered states are 
highly directionally entangled, in analogy with the effect of an optical beam-splitter. 
However, in contrast to the classical beam splitter, the high correlation seen here is induced solely by non-linearities. 
To demonstrate that the high directional correlations indeed stem from non-linearities, in \figref{comparison}(a) 
we also show the case where the non-linear two-photon interaction term $b_{12}(p,p')$ has been artificially set to $0$ (dashed curves). For an uncorrelated two-photon input pulse which is resonant with the emitter and which has a symmetric spectral wavefunction amplitude, $|\xi(-k)|^2=|\xi(k)|^2$, as is the case here, \eqref{eq_P11} reduces to
\begin{align}
&P_{11}=a(1-a), \nn \\
&\text{with  } a = \dintegral{}{p}{-\infty}{\infty}\frac{(\tilde{\Gamma}/2)^2}{p^2+(\tilde{\Gamma}/2)^2}|\xi(p)|^2,
\end{align}
showing that $P_{11}$ maximally attains the value $1/4$, occurs when $a=1/2$. Thus, $F_\text{prob}=2P_{11}$ never exceeds 1/2, as confirmed in  \figref{comparison}(a), which indicates that no directional entanglement is present in the scattered state~\cite{Kokbook}. For $b_{12}(p,p')=0$, the emitter behaves as a linear component (e.g. a lossless optical cavity) and 
cannot mediate interactions between the two photons. As such, the scattering process is 
determined entirely by interference effects, which, unlike an optical beam-splitter, cannot create entanglement in 
this system when the pulses are resonant with the emitter.

A more direct analogy with a 50/50 beam splitter can be obtained by detuning the input pulses by half the 
emitter linewidth, $\Delta = \Gamma/(2v_g)$. For this value of the detuning, a monochromatic single-photon pulse 
will be reflected/transmitted with 50\% probability 
(whereas at $\Delta=0$ a monochromatic single-photon pulse is fully reflected). 
\figref{comparison}(b) shows $F_{\mathrm{prob}}$ for $\Delta = \Gamma/(2v_g)$, and we 
confirm that $F_{\mathrm{prob}}\to 1$ as $\sigma\to 0$ as expected. 
The change in the fidelity due to the non-linearities now becomes smaller than in the resonant case, 
as the interaction between the pulse and the emitter 
is less efficient off resonance. Interestingly, in this case, for small $\sigma$, the non-linear interaction 
actually deteriorates the beam splitting effect, since now the directional entanglement can be generated 
by interference effects only.

For a Lorentzian input 
analytic expressions for $P_{11,\text{Lor}}=P_{22,\text{Lor}}$ may be derived. We find 
\begin{equation}
P_{11,\text{Lor}}=\frac{3\tilde{\Gamma}\sigma(3\sigma\!+\!\tilde{\Gamma})(\sigma\!+\!\tilde{\Gamma})+4\Delta^2\tilde{\Gamma}(\sigma\!+\!2\tilde{\Gamma}) }{ \left[(3\sigma+\tilde{\Gamma})^2+4\Delta^2\right]\left[(\sigma+\tilde{\Gamma})^2+4\Delta^2\right]}
\end{equation}
\begin{figure*}[t]
\includegraphics[scale=0.6]{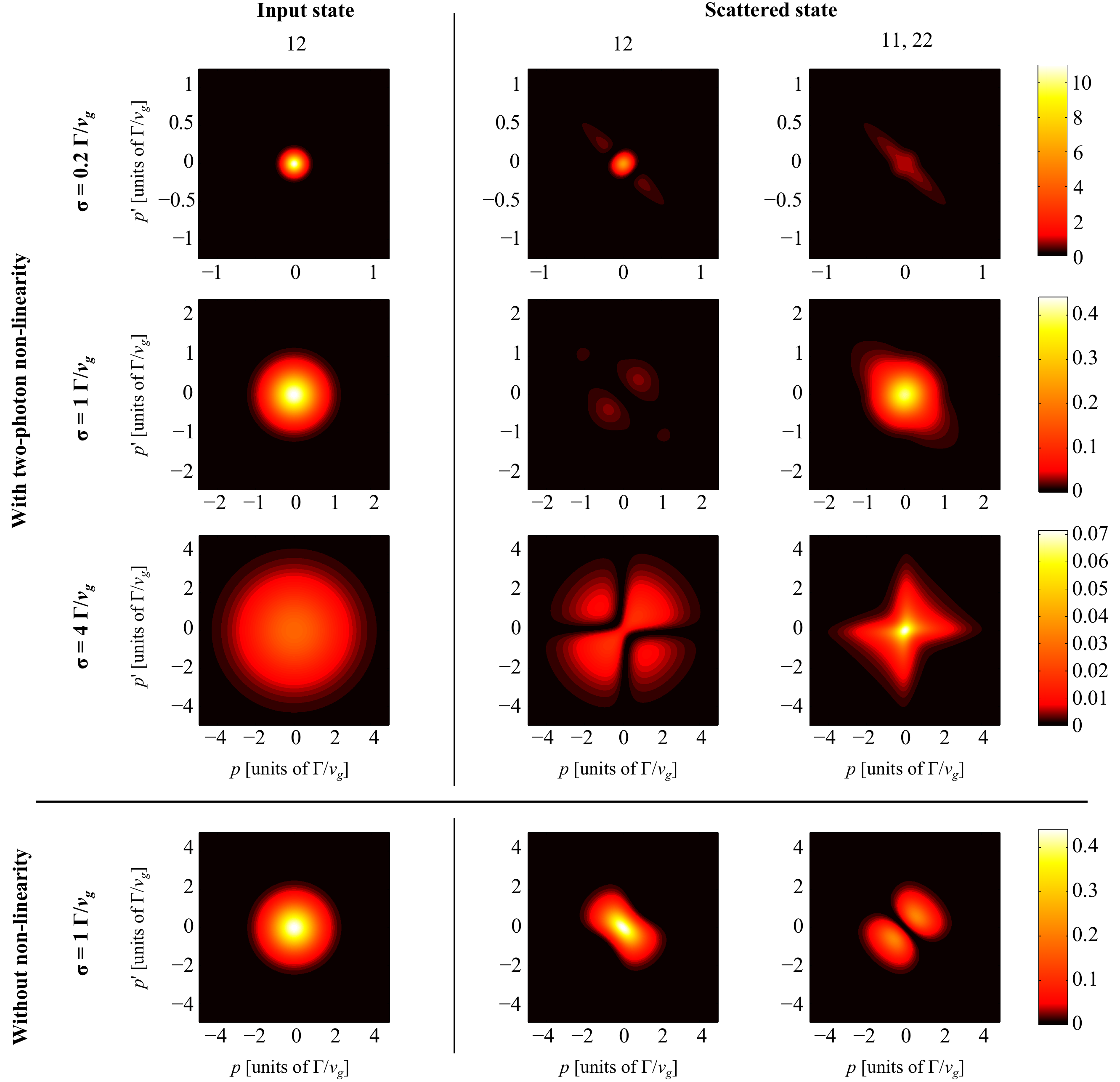}
\caption{\label{fig:examples_wf} Intensity spectrum of an incoming Gaussian two-photon state for two 
counter-propagating photons (left column), and the resulting two-photon intensity spectra for 
the scattered state (middle and right columns) for photons scattering in different directions, 12, and where both photons propagate in the 
same direction, 11 (identical to 22), with $\Delta=0$. The spectral width of the input pulses is varied: $\sigma=0.2\,\tilde{\Gamma}$ (1st row), 
 $\sigma=1\,\tilde{\Gamma}$ (2nd row), and $\sigma=4\,\tilde{\Gamma}$ (3rd row). The intensity spectra are also shown for scattering with the non-linearity turned off, $b_{12}(p,p')=0$, using $\sigma=1\,\tilde{\Gamma}$ (4th row).}
\end{figure*}
with $\tilde{\Gamma}=\Gamma/v_g$. 
The maximum value on resonance ($\Delta=0$) is obtained for 
$\sigma/\tilde{\Gamma}=3^{-1/2}\approx 0.57$, at which point $P_{11,\text{Lor}}\approx 0.40$ 
(and $F_{\mathrm{prob}}=2P_{11,\text{Lor}}\approx 0.8$), in agreement with \figref{comparison}(a). 
In comparison, with no non-linear terms, $b_{12}(p,p')=0$, the scattering probability becomes 
\beqa
P^\text{one}_{11,\text{Lor}}=\frac{ \sigma\tilde{\Gamma}(\sigma+\tilde{\Gamma})^2+4\Delta^2\tilde{\Gamma}(\sigma+2\tilde{\Gamma}) }{ \left[(\sigma+\tilde{\Gamma})^2+4\Delta^2 \right]^2 }
\eeqa
\noindent which has an on-resonance maximum for $\sigma=\tilde{\Gamma}$, 
giving $F_{\mathrm{prob}}=2P^\text{one}_{RR,\text{Lor}}=1/2$. 
Thus, for the Lorentzian pulse, non-linearities increase the scattering probability by a factor of 
$P_{11,\text{Lor}}/P^\text{one}_{11,\text{Lor}}=1+2/(1+3\sigma/\tilde{\Gamma})$ on resonance. 
For $\sigma\rightarrow \infty$, the interaction with the emitter becomes infinitely weak and no enhancement is present. 
In the opposite limit of $\sigma\rightarrow 0$, the enhancement factor is 3. 

\subsection{Scattering Fidelities}

As discussed above, if the scatterer is to be implemented in a larger optical circuit, in addition to 
considering in which \emph{direction} the photons scatter, the \emph{amplitude} and \emph{phase} of the 
different frequency components may also be important. In such a case, 
$F_\text{prob}$ is no longer a sufficient fidelity measure, since it contains only directional information. 
As illustrated in \figref{examples_wf}, where we plot the intensity spectrum of a scattered Gaussian wavepacket, 
the spectra of the scattered pulses change significantly during the scattering process. For input pulses with a 
narrow spectral linewidth compared to the emitter (1st row), the pulse power at the emitter position remains low 
due to the corresponding broad spatial profiles of the pulses. In that case, the non-linearity is only weakly 
addressed, and the individual photons are predominately reflected. A weak non-linearity-induced four-wave mixing 
process is signified by the appearance of diagonal features, as one photon achieves a larger energy, while  
the energy of the other decreases. When the pulse and emitter linewidths are 
comparable (2nd row), the predicted strong directional correlation is induced~\cite{Nysteen14}, 
with the pulse profile being almost preserved. For spectrally broad pulses (3rd row), only the near-resonant part of 
the spectrum interacts with the emitter. We see that the spectral components at the emitter 
frequency are absent from the transmitted pulse since these have been reflected 
without significant two-photon effects. 

Interestingly, the fact that the pulse spectrum is almost perfectly preserved when the pulse and emitter linewidths are comparable (2nd row) can be attributed to non-linearities. This can be seen in the 4th row, where we again show the initial and scattered spectra for the case $\sigma=\tilde{\Gamma}$, but where we have artificially set the non-linear term equal to zero, $b_{12}=0$. By comparison with the 2nd row, we can clearly see that 
the non-linearities not only give rise to the directional entanglement, but also suppress changes to the spectral shape.

\begin{figure}
\includegraphics[width=0.45\textwidth]{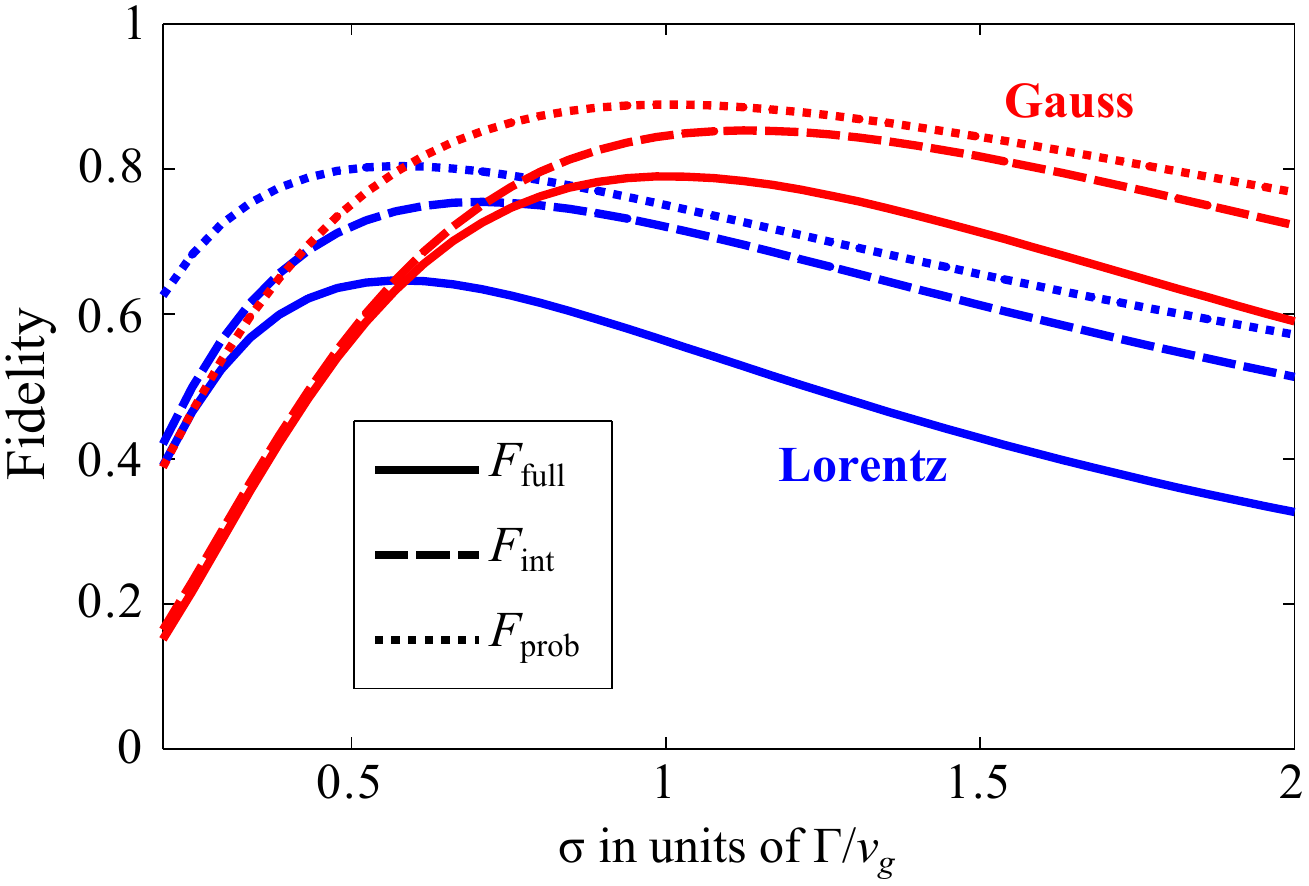}
\caption{\label{fig:two_fids} Fidelities from Eqs. (\ref{eq:two_2fid1})-(\ref{eq:two_2fid3}) plotted for varying 
width of the input pulses, shown both for Gaussian and Lorentzian inputs on resonance with the emitter.}
\end{figure}

To quantify both the spectral and phase deviations between the scattered and the desired state, 
all three fidelities defined in Eqs.~(\ref{eq:two_2fid1})-(\ref{eq:two_2fid3}) are shown in \figref{two_fids}. 
By comparing $F_\text{prob}$ to $F_\text{int}$, i.e. taking into account the difference in the spectra  
of the scattered and desired pulse (but not the phase), we see that the fidelity becomes lower, and most significantly so for 
pulses with a small spectral linewidth. This can be understood from \figref{examples_wf}, 
where we see that the scattered 
wavepacket for the spectrally narrow input (1st row) is clearly influenced by strong four-wave mixing effects. 
For pulses with larger widths, these effects are weaker, since a larger fraction of frequency 
components are detuned from the emitter transition and therefore interact only weakly. 

Scattering-induced phase differences across the pulses may be examined by 
comparison of $F_\text{int}$ and $F_\text{full}$. As illustrated in \figref{two_fids}, these fidelities are 
almost equal for spectrally narrow pulses, whereas significant deviations are seen for spectrally broad pulses. 
This may be explained by considering the simpler single-photon scattering case. 
From \eqref{one_reftrans}, we see that a resonant, monochromatic pulse will be reflected with a phase shift of $\pi$, 
whereas spectrally broader pulses attain a phase shift from $\pi/2$ to $3\pi/2$ across the pulse spectrum. 
Thus, spectrally broad pulses experience larger decreases in the fidelity due to phase mismatching with our given desired state. 

\begin{figure}[b]
\includegraphics[scale=0.6]{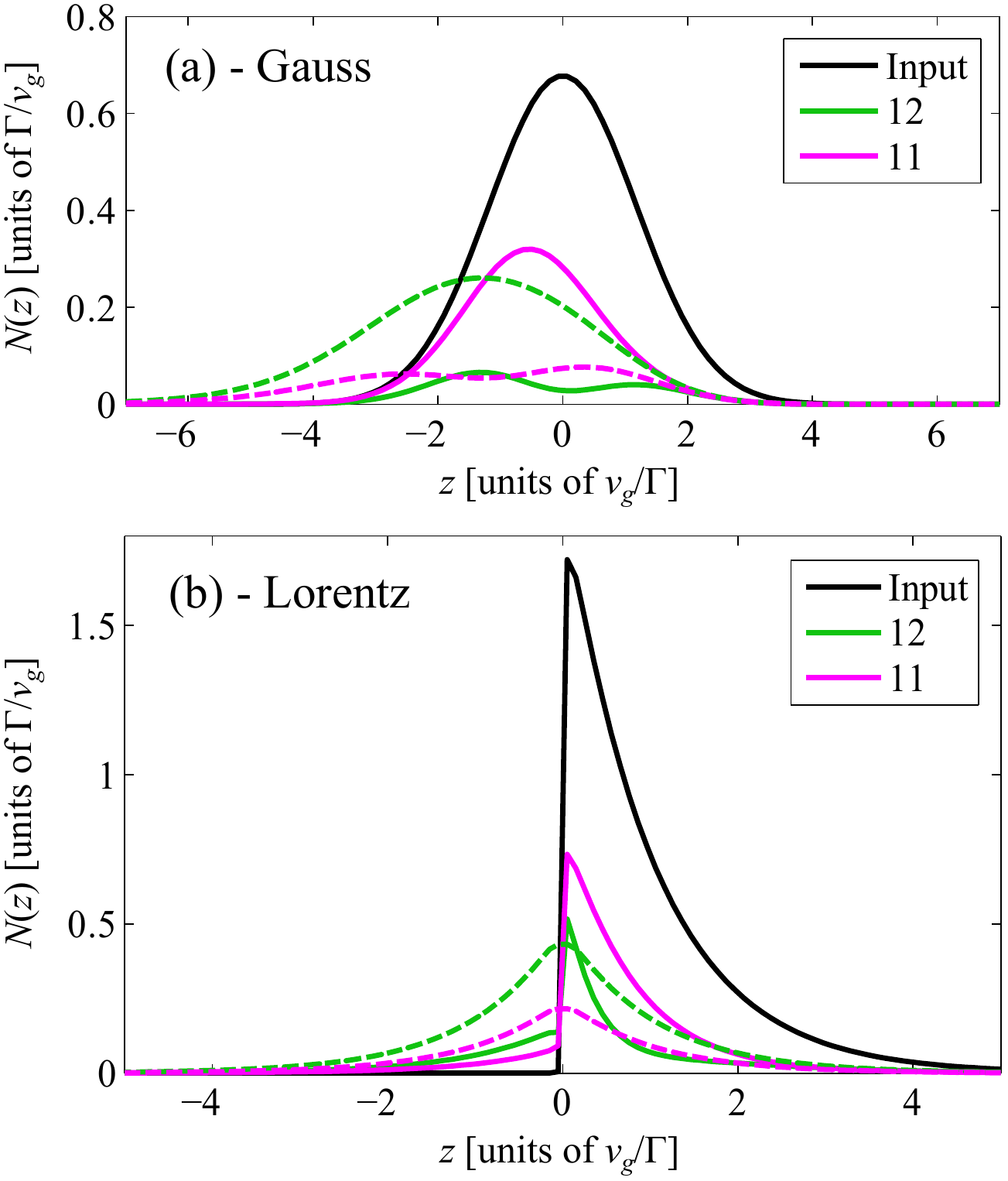}
\caption{\label{fig:two_Nz} The photon density in the moving frame, $N(z)$, calculated for the input pulse and for 
photons in the same (11 identical to 22) or in different modes after the scattering (12) for resonant input pulses with 
$\sigma=1 \,\tilde{\Gamma}$ and $\Delta=0$. The largest values of $z$ correspond to the front part of the pulse, and the solid (dashed) lines include (do not include) the 
non-linear two-photon interaction, $b_{12}(p,p')$. (a) Gaussian wavepacket (b) Lorentzian wavepacket.}
\end{figure}

As the phase changes correspond to modifications to the spatial profile of the pulse, we specifically consider how the spatial pulse profile 
(here analogous to the temporal shape) is changed during the scattering process. To clearly illustrate the effect of the non-linear scattering on the spatial profile, we evaluate the photon density at a specific 
point of the photon wavepacket in the rotating frame, defined as 
\beqa
N(z)= _{t\rightarrow\infty}\bra{\beta}a^\dagger(z)a(z)\ket{\beta}_{t\rightarrow\infty}
\eeqa
where $a(z)=(2\pi)^{-1/2}\dintegral{}{k}{-\infty}{\infty}a(k)\exp[\ii k z]$  
is the annihilation operator for an excitation at a position $z$ in a frame rotating with the pulses. 
For Gaussian and Lorentzian input pulses, the photon density is plotted in \figref{two_Nz}. For both pulse shapes, we see that a delay occurs due to interaction with the emitter, and furthermore the non-linearity improves the similarity between the scattered and incoming field. The Gaussian pulse is observed 
to preserve its spatial symmetry, as compared to the Lorentzian input pulse. This is due to the fact that the part of the 
photon pulses which is absorbed by the emitter is re-emitted with an exponential shape that is spatially reversed 
compared to the input pulse, which explains why $F_\text{prob}$ deviates significantly from $F_\text{int}$ for large spectral linewidths in \figref{two_fids}.

\section{Conclusion}
We have analytically demonstrated that the non-linearity of a two-level-emitter can induce strong pulse-dependent directional correlations (entanglement) in the scattered state of two initially counter propagating photons. These correlations are maximized for photons with spectral widths comparable to that of the emitter, and also depend on the specific spectral shape of the photons. Furthermore, we have investigated how the spectra and phase of the photons are affected by the scattering process, and introduced different fidelity measures to quantify the similarity of the scattered and input photons. Interestingly, for photons with spectral widths comparable to the emitter linewidth, where the directional correlations are maximized, the non-linearity of the emitter acts to suppress changes in the spectra and phase of the photons. As such, even when taking all properties of the scattered state into account, a comparison to perfect directionally entangled photons with preserved spectra and phases gives fidelities as high as $\sim 80\%$ for Gaussian pulse shapes. A comparison of our fidelity measures indicates that when engineering photonic gate structures and other functionalities using two-level-emitters, it is important to also consider spectral and phase changes when determining the efficiency and scalability of non-linear photonic devices.

\section*{Acknowledgements}
This work was supported by Villum Fonden via the Centre
of Excellence ``NATEC'', the Danish Council for Independent Research (FTP 10-093651) and by the European Metrology Research
Programme (EMRP) via the project SIQUTE (Contract
No. EXL02). The EMRP is jointly funded by the EMRP participating countries within EURAMET and the European Union.


%
%

\bibliographystyle{apsrev}

\end{document}